\documentclass[twocolumn]{aastex62}

\usepackage{graphicx}
\usepackage{graphbox}
\usepackage{tikz}
\usepackage{float}
\usepackage{multirow}
\usepackage{color}
\usepackage[colorinlistoftodos]{todonotes}

\usepackage[italicdiff]{physics}

\begin{document}
	\title{TauREx3 PhaseCurve: A 1.5D model for phase curve description.}
	
	\correspondingauthor{Q. Changeat}
	\email{quentin.changeat.18@ucl.ac.uk}
	\author[0000-0001-6516-4493]{Q. Changeat}
	\affil{Department of Physics and Astronomy \\
		University College London \\
		Gower Street,WC1E 6BT London, United Kingdom}
	\author[0000-0002-4205-5267]{A. Al-Refaie}
	\affil{Department of Physics and Astronomy \\
		University College London \\
		Gower Street,WC1E 6BT London, United Kingdom}

\submitjournal{The Astrophysical Journal}
\accepted{09 June 2020}

%%%%%%%%%%%%%%%%%%%%%%%%%%%%%%%%%%%%%%%%%%%%%%%%%%%%%%%%%%%%%%%%%%%%%%%%%%%%%%%%
\begin{abstract}

In the recent years, retrieval analysis of exoplanet atmospheres have been very successful, providing deep insights on the composition and the temperature structure of these worlds via the transit and eclipse methods. Analysis of spectral phase curve observations, which in theory provides even more information, are still limited to a few planets. In the next decade, new facilities such as NASA-JWST and ESA-Ariel will revolutionise the field of exoplanet atmospheres and we expect that a significant time will be spent on spectral phase curve observations. Most current models are still limited in their analysis of phase curve data as they do not consider the planet atmosphere as a whole or they require large computational resources. In this paper we present a semi-analytical model that will allow to compute exoplanet emission spectra at different phase angles. Our model provides a way to simulate a large number of observations while being only about 4 times slower than the traditional forward model for plane parallel primary eclipse. This model, which is based on the newly developed TauREx\,3 \citep{al-refaie_taurex3} framework, will be further developed to allow for phase curve atmospheric retrievals.

\end{abstract}

%%%%%%%%%%%%%%%%%%%%%%%%%%%%%%%%%%%%%%%%%%%%%%%%%%%%%%%%%%%%%%%%%%%%%%%%%%%%%%%%
\section{INTRODUCTION}

The field of exoplanetary atmospheres has seen a rapid development of novel methods and techniques. Some of the more recent breakthroughs include spatial scanning methods using WFC3 \citep{McCullough_scanning_HST}, automated data reduction pipelines \citep{Tsiaras_2016_iraclisHD209} and retrievals using Bayesian sampling methods \citep{Waldmann_taurex1, Waldmann_taurex2, Irwin_nemesis, Line_chimera, Ormel_arcis, harrington_bart, Mollire_petitrad, Kitzmann, MacDonald_hd209, Gandhi_retrieval, benneke_retrieval, Zhang_platon, cubillos_pirat}. 
Most current retrieval analysis rely on specific geometric configurations such as transits, when the planet passes in front of its host star: \citep{Tsiaras_pop_study_trans, sing_exoplanets}, or eclipses, when the planet passes behind the star: \citep{Evans_Wasp121b_spectrum_em, Haynes_Wasp33b_spectrum_em}. 
These configurations give insight into the day-night interface and day-side atmosphere respectively. The limitation on geometry stems from the generally low signal-to-noise of current instrumentation and sparsity of observation facilities hindering multiple observations of the same target at different configurations. Due to the relatively low information content in current available spectra, the use of 1D models is well justified. There has been growing interest in phase-curves, spectra from a range of geometries or phase angles, spurred on from the handful of targets combining the good conditions to produce them. Phase curves do not benefit from a particular configuration (as opposed to transit and eclipse observations). With the next generation of space telescopes (\citealp[NASA-JWST:][]{Greene_2016_jwst, Bean_JWST}; \citealp[ESA-Ariel:][]{Tinetti_ariel}), planetary atmospheres will be studied extensively and phase curve observations will be obtained for a larger number of targets.
Analysis of current phase curve datasets have revealed important physical phenomena including: shifts of the dayside hot-spot, high day-night contrasts and other effects from atmospheric dynamics \citep{stevenson_w43_1,stevenson_w43_2, De_wit_hd189_hotspot, Zellem_hd209_phase, carone_equatorial}. However the current standard approach of retrieving spectra as individual, independent measurements does not exploit the spatial information provided. In that context, it is important to study the feasibility of accumulating \citep{Yip_lightcurve} such observations and to develop the necessary tools to ensure an optimal and complete extraction of information. Recent studies \citep{Feng_2016,caldas_3deffects,Irwin_w43b_phase,2020_pluriel,Taylor_2020,MacDonald_2020} highlighted the impact of 3-dimensional effects on exoplanet spectra and the importance of combining the different phases under a common atmospheric model, abandoning the 1D model assumption. In \cite{Irwin_w43b_phase}, the authors highlighted the difficulties linked to the high computing requirements of their model, which translated into limitations in their retrieval sampling method to optimal estimation. Here we propose an alternative model to describe phase curve scenarios, in which the geometry is computed analytically. An independent, similar approach is also described in \cite{feng2020_2d}.
Our model is implemented in latest version of TauREx\,3 \citep{al-refaie_taurex3} providing increased computational efficiency and high flexibility. In the first section, we describe the calculation of the phase curve model. Then we produce an example based on WASP-43\,b to illustrate the possibilities of the model and provide some comparison with the literature. Finally, in the discussion section we benchmark the performances and the limitations of our model.

\section{PHASE CURVE MODEL}

\subsection{Structure of the model}

We build our phase curve model using the latest version of TauREx\,3 \citep{al-refaie_taurex3}, which is the most recent rework of TauREx \citep{Waldmann_taurex2, Waldmann_taurex1}.

For this 1.5D phase curve model, we assume that the planet consists of 3 distinct regions: a day side, a terminator region and a night side. Each region is characterised by its own emission model (respectively $E_d$ for the day side, $E_t$ for the terminator and $E_n$ for the night side), built from the pre-existing TauREx 3 eclipse model \citep{Waldmann_taurex1}. For each phase, the fractional contribution from each regions integration point must sum to unity. We also assign a transmission model $T$ to the terminator region as to include transit spectra in the model (corresponding to phase around zero).

The choice and behaviour of each atmospheric parameter is chosen freely by the user. Each region can be completely decoupled; with each behaving as three/four separate forward models. Completely coupled; where all parameters are shared between regions or a mixture of the two (e.g coupling the terminator and night-side whilst leaving the day side free). This applies to each individual atmospheric parameter for each region providing a high degree of flexibility in model choice.
For instance, a possible configuration could be to couple the same Guillot \citep{Guillot_TP_model} temperature profile with the terminator and night side and a more flexible 3-point profile \cite{al-refaie_taurex3} in the day side whilst using an equilbrium chemistry model for the day and a coupled free-type on the terminator and night with each region having their own treatment of clouds. Retrievals for decoupled parameters have the \texttt{day\_}, \texttt{term\_} and \texttt{night\_} prefix (e.g \texttt{day\_T} for isothermal temperature in the day side). Certain parameters such as the planet radius $R_p$ and the planet mass $M_p$ are always coupled.

\subsection{Basic Transmission and Emission models}

As previously stated, the transmission and emission models are built from the native ones in TauREx \citep{Waldmann_taurex1, Waldmann_taurex2, al-refaie_taurex3}. For completeness, we have repeated the equations used. In the transmission case. The observed signal $\Delta_{\lambda}$ is:
\begin{equation}
    \Delta_{\lambda} = \left(\frac{R_p}{R_s} \right)^2 + \frac{2}{R_s^2} \int_0^{z_{max}}(R_p+z)(1-e^{-\tau_{\lambda}(z)}) dz,
\end{equation}
where $R_s$ is the radius of the star and $\tau_{\lambda}(z)$ is the wavelength dependant optical depth as a function of altitude $z$. \\ 

In the emission case, the observed signal is described by the following equation:

\begin{equation}
    \label{equation_secondary}
    \frac{F_p}{F_s} = \left(\frac{R_p}{R_s}\right)^2 \times \frac{I_{\lambda}(\tau = 0)}{I_s},
\end{equation}

where $I_s$ is the wavelength dependence stellar intensity and  $I_{\lambda}(\tau = 0)$ is the intensity at the top of the exoplanet atmosphere. We note $\theta$ the viewing angle and $\mu = $cos$(\theta)$. $I_{\lambda}(\tau = 0)$ is defined as:
\begin{equation}\label{eq:base_em}
    I_{\lambda}(\tau = 0) = B_{\lambda}(T_s) e^{-\tau_s/\mu} + \int_0^{\tau_s}B_{\lambda}(T_{\tau}) e^{-\tau/\mu} \frac{d\tau}{\mu},
\end{equation}
where $B_{\lambda}(T)$ is the Plank function at a given temperature T.

Now the total flux is an integral of the projected planet disk surface. We use the Gaussian quadrature method to perform the integral over the viewing angles and denote $\omega_i$ the quadrature weights and $\mu_i$ the quadrature points indexed by i. $\mu_i$ corresponds to the integration of the circle at radius $\mu_i = $cos$(\theta_i)$. The total number of quadrature points is $N_G$.

Therefore, the calculation of $I_{\lambda}$ is split into $N_G$ calculations of $I_{\lambda, i}$ corresponding to the viewing angle $\theta_i$ and the total flux is given by:

\begin{equation}
    \label{equationGauss}
    I_{\lambda}(\tau = 0) = 2 \pi \sum_i^{N_G} I_{\lambda, i} \times \omega_i \times \mu_i 
\end{equation}

\subsection{Phase dependent emission model}

For the phase dependent emission, we combine the contributions of the three regions: day, terminator and night.

For a given phase angle $\Phi$, where $\Phi$ represents the angle between the star-planet and star-observer axes, we use the same projection onto the 2d disk to calculate the emission.

We know the terminator must pass through the three points (cos$(\Phi)$, 0), (0,1) and (0,-1) defined on the (x,y) orthonormal basis. It must also be equivalent to a circle at phase 180 and be symmetric along the y axis for phase 90. To match these conditions, we assume that the terminator projection takes the form of an arc circle passing through the three previously mentioned points. Then the terminator region is defined by the arc circles of same centre but with smaller/larger radius using $K_{\pm}$, where $K_{+}$ and $K_{-}$ are the projected distances from the centre of the terminator to the boundaries. $K_{\pm}$ therefore describes the size of the terminator region on the 2d disk and can be related to the terminator spherical angle size $\theta_K$ by:
\begin{equation}
    K_{\pm} = |\mathrm{cos}(\Phi) - \mathrm{cos}(\Phi \pm \theta_K)|
\end{equation}

We note $\mu = $cos$(\theta)$ the angle between the planet normal and the planet-observer axis, so sin$(\theta)$ is the radius of the integration disk for each Gaussian point. Figure \ref{fig:geometry} represents the geometry of the problem, where we show the 3 regions as well as an example of integration circle of radius sin$(\theta)$:

%\onecolumngrid

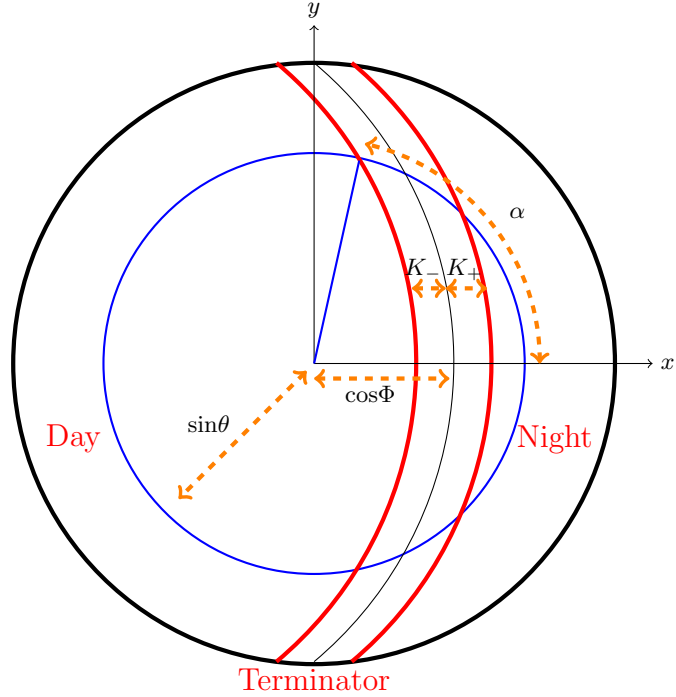
\begin{figure*}
\begin{center}
\begin{tikzpicture}

\draw [ultra thick] (0,0) circle (4);
\draw [color=blue, thick] (0,0) circle (2.8);
\draw[] (-0,4) arc (50:-50:5.2cm);
\draw[color=red, ultra thick] (-0.5,4) arc (50:-50:5.2cm);
\draw[color=red, ultra thick] (+0.5,4) arc (50:-50:5.2cm);
\draw [->] (0,0) -- (0,4.5);
\draw [->] (0,0) -- (4.5,0);
\node (a) at (4.7, 0) {$x$};
\node (b) at (0, 4.7) {$y$};

\draw [dashed, <->,  color=orange, ultra thick] (-0.1,-0.1) -- (-1.8,-1.8);
\node (c) at (-1.4, -0.8) {sin$\theta$};
\draw [dashed, <->, color=orange, ultra thick] (0,-0.2) -- (1.8,-0.2);
\node (d) at (0.75, -0.4) {cos$\Phi$};

\draw [dashed, <->,  color=orange, ultra thick] (1.3,1) -- (1.75,1);
\node (e) at (1.45, 1.25) {$K_{-}$};
\draw [dashed, <->,  color=orange, ultra thick] (1.75,1) -- (2.3,1);
\node (f) at (2, 1.25) {$K_{+}$};

\draw [-, color=blue, thick] (0,0) -- (0.6,2.7);
%\draw[dashed] (0.5,0) arc (0:75:0.5cm);
%\node (f) at (0.6, 0.5) {$\alpha$};

\draw[dashed, <->,  color=orange, ultra thick] (3,0) arc (0:77:3cm);
\node (g) at (2.7, 2) {$\alpha$};

\node[text=red, thick] (h) at (-3.2, -1) {\large Day};
\node[text=red, thick] (i) at (3.2, -1) {\large Night};
\node[text=red, thick] (j) at (0, -4.2) {\large Terminator};

\end{tikzpicture}
\end{center}
\caption{Illustration of our simplified phase geometry. The three regions are represented (the black circle represents the planet boundary and the red separations are for the terminator) as well as the necessary parameters to constrain their geometry. We also show an example of line integral (blue circle) at distance sin$(\theta)$, corresponding to the Gaussian point $\mu = $cos$(\theta)$. The parameter we are looking to constrain is $\alpha$ as a function of $\mu$, $\Phi$ and K since it represents the coefficients of the different regions for this Gaussian point $\mu$. } \label{fig:geometry}
\end{figure*}

%\twocolumngrid

Now the objective is to calculate, for each Gaussian point, the contribution of the different regions. We define $C^d$ as the contribution from region $E_d$, $C^t$ the contribution from region $E_t$ and $C^n$ the contribution from region $E_n$. This is equivalent to calculating the angles from the x axis to the intersection of the terminator boundaries and the integration circle. We consider the planet of size 1 in arbitrary units and perform this computation analytically (see Appendix 1 for the detailed derivation). For a given phase $\Phi$, a Gaussian point $\mu$ and a terminator size K we find that the angle $\alpha$ from the x axis to the point of intersection between the integration disk and the terminator region is given by:

\onecolumngrid
\noindent\rule{18cm}{0.4pt}
\begin{equation}\label{eq:phase_eq}
    \alpha(\Phi,\mu, K_{\pm}) = \mathrm{arccos} \left( \frac{\mathrm{cos}(\Phi)}{ (1-\mathrm{cos}^2(\Phi))\sqrt{1-\mu^2}}\left( \mu^2 + K_{\pm}^2 \pm 2 K_{\pm} \sqrt{1 +\frac{(\mathrm{cos}^2(\Phi)-1)^2}{4 \mathrm{cos}^2(\Phi)}} \right) \right),
\end{equation} 

For the case $\Phi = \pi/2$, we use:

\begin{equation}
    \alpha(\pi/2,\mu, K_{\pm}) = \mathrm{arccos} \left( \frac{ \mathrm{sin}(\theta_K)}{\sqrt{1-\mu^2}} \right),
\end{equation}
\noindent\rule{18cm}{0.4pt}
\twocolumngrid

For each Gaussian point, we perform the calculation of the angle for the terminator boundaries $K_{-}$ and $K_{+}$ that we denote respectively $\alpha_-$ and $\alpha_+$. The angles $\alpha_{-}$ and $\alpha_{+}$ from Equation \ref{eq:phase_eq} are ill defined when the integration circle does not intersect with the terminator boundaries. These cases need to be handled individually, giving
rise to 5 distinct cases for the $C$ coefficients: \\ \\

$\bullet$ $\alpha_-$ and $\alpha_+$ are not defined and $cos\Phi - K > sin \theta$: In this case, the integration circle is entirely inside the region $E_d$ so the coefficients are: \\
    - $C^n = 0$ \\
    - $C^t = 0$ \\
    - $C^d = 1$\\ \\ 
    
$\bullet$ $\alpha_-$ and $\alpha_+$ are not defined and $cos\phi - K < sin \theta$ and $cos\Phi + K > sin \theta$: In this case, the integration circle is entirely inside the region $E_t$ so the coefficients are: \\
    - $C^n = 0$ \\
    - $C^t = 1$ \\
    - $C^d = 0$\\ \\ 
    
$\bullet$ $\alpha_+$ is not defined: In this case, the integration circle is shared by the region $E_d$ and the region $E_t$ so the coefficients are: \\
    - $C^n = 0$ \\
    - $C^t = 2 \alpha_-$ \\
    - $C^d = 1 - 2 \alpha_-$\\ \\ 

$\bullet$ $\alpha_-$ and $\alpha_+$ are defined: In this case, the integration circle cuts all three regions so the coefficients are: \\
    - $C^n = 2 \alpha_+$ \\
    - $C^t = 2 ( \alpha_- - \alpha_+ )$ \\
    - $C^d = 1 - 2 \alpha_-$\\ \\ 
    
$\bullet$ $\Phi = \pi/2 $: This is a particular case, if $\alpha_+$ is not defined, we use $\alpha_+ = 0$. Then: \\
    - $C^n = 2 \alpha_+$ \\
    - $C^t = 1 - 4 \alpha_+$ \\
    - $C^d = 2 \alpha_+$ \\ \\

This simple analytic form allows for the precalculation of coefficients. The final emission at a given phase is given by modifying equation \ref{equationGauss} to include the different contributions:

\begin{equation}
    I_{\lambda}(\tau = 0) = 2 \pi  \sum_i^{N_G} \left( I_{\lambda, i}^d C^d_i + I_{\lambda, i}^t C^t_i + I_{\lambda, i}^n C^n_i \right) \times \omega_i \times \mu_i,
\end{equation}
where $I_{\lambda, i}^d$, $I_{\lambda, i}^t$ and $I_{\lambda, i}^n$ are the day, terminator and night intensities at the top of the atmosphere for the Gaussian point $\mu_i$. $C^d_i$, $C^t_i$ and $C^n_i$ are the contribution of the regions D, T and N for the Gaussian point $\mu_i$. \\

Now this can be integrated back in equation \ref{equation_secondary}, taking into account for the contribution of the 3 different regions as a function of phase. 

\begin{center}
\begin{figure*}
\centering
    \includegraphics[width=1\textwidth]{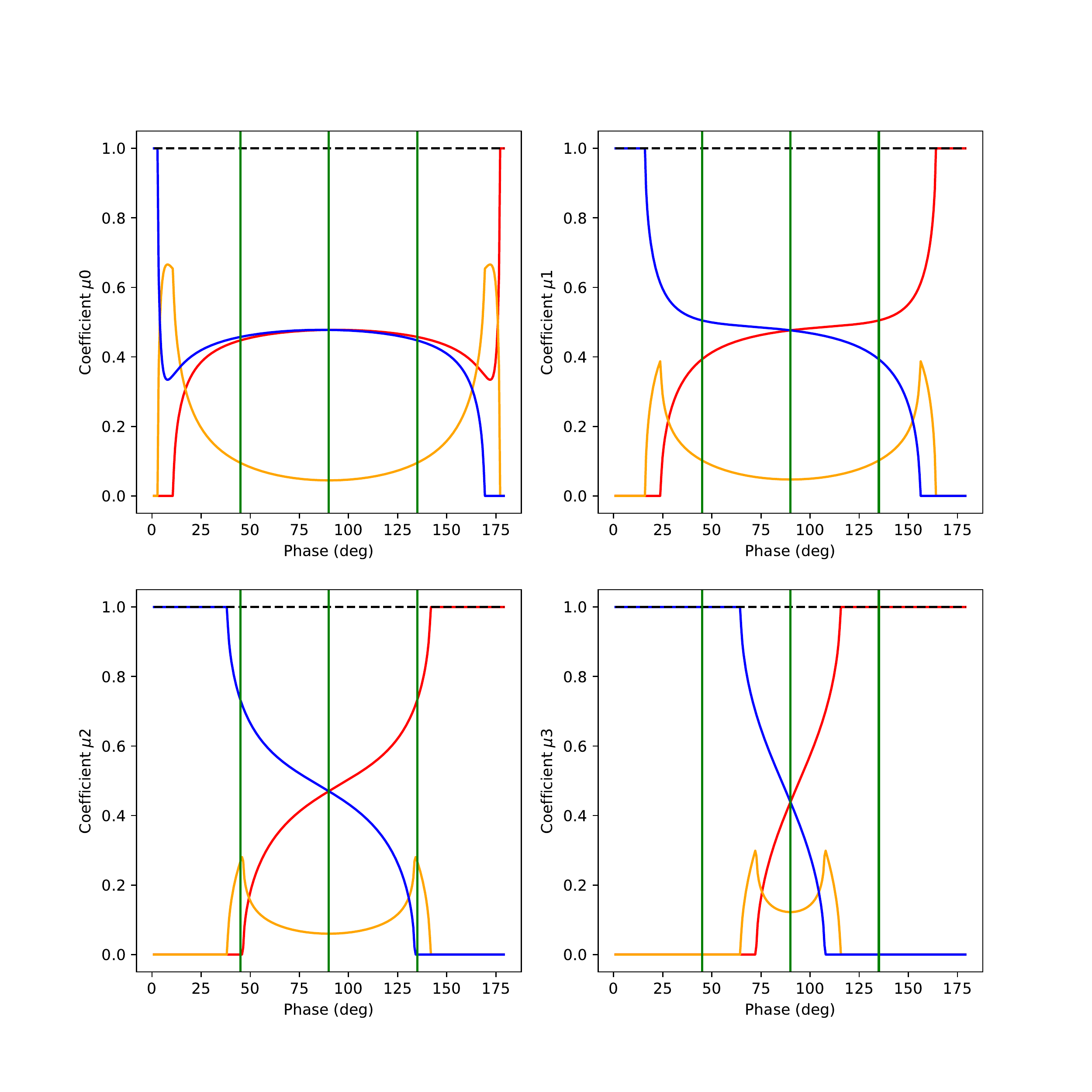}
\caption{Value of the phase coefficients for a model with 4 quadrature points arranged from the exterior ($\mu0$) to the centre ($\mu3$). Red: Day side coefficients; Orange: Terminator side coefficients; Blue: Night side coefficients. The Green vertical lines highlight the coefficients at phases 45 degrees, 90 degrees and 135 degrees.}
\label{fig:phase_coeffs}
\end{figure*}
\end{center}

We show in  Figure \ref{fig:phase_coeffs} the evolution of the phase coefficients as a function of the phase in the case where the number of Gaussian quadrature points is $N_G = 4$ (e.g: $\mu 0 = 0.1834346$; $\mu 1 = 0.5255324$; $\mu 2 = 0.7966665$; $\mu 3 = 0.9602899$).  

 Equation \ref{eq:phase_eq} can be used directly for planets in circular orbits (eccentricity e=0) and with no inclination (I = 90) since the phase angle $\Phi$ is linear with time in this case. In other cases, a change of variable is required to calculate the phase angle $\Phi$ as a function of time ($\Phi(t)$). This calculation can be performed using Kepler's laws (see Appendix 2 for the derivation of $\Phi(t)$ in the case of tidally locked planets) and allows us to generalise Equation \ref{eq:phase_eq}. An example of a planet in elliptical orbit and the corresponding evolution of the phase angle $\Phi$ can be found in Appendix 3. In addition, if the planet is not tidally locked but in synchronous resonance, an additional correction can be introduced to calculate the phase angle corresponding to the viewed face \citep{sertorio2001constraints}. In this case, $\Phi(t)$ transforms to:
\begin{equation}
    \Phi_{sync}(t) = \frac{T}{T_d}\Phi_{tid}(t) + \Phi_0,
\end{equation}
where $\Phi(t)_{sync}$ is the corrected phase angle for synchronous orbits, $\Phi_{tid}(t)$ is the phase angle calculated in Appendix 2 for a tidally locked planets. T is the orbital period and T$_d$ is the period corresponding to a planet revolution around its spin axis and $\Phi_0$ is an eventual initial angular offset.

%\onecolumngrid
%\twocolumngrid

\section{Forward model example}

In this section we present an example of a phase-curve forward model for the hot-Jupiter WASP-43\,b. Its phase-curve has been extensively studied in \citep{stevenson_w43_1, stevenson_w43_2, Irwin_w43b_phase, morello_w43_phase}. It possesses a large day-nigh contrast, and a sharp transition at the terminator. We use this example to illustrate our phase curve model but its interpretation is beyond the scope of this article. While the models are very different, we take inspiration from the retrieval analysis of \cite{stevenson_w43_2} for our input parameters. Our forward model includes the molecular cross sections from the Exomol project \citep{Tennyson_exomol}, HITEMP \citep{rothman} and HITRAN \citep{gordon}: H$_2$O \citep{barton_h2o, polyansky_h2o}, CH$_4$ \citep{exomol_ch4, hill_xsec} and CO \citep{li_co_2015}. These opacities are sampled at a resolution of $R=15000$ from 0.3 $\mu$m to 50 $\mu$m. We add collision induced absorption for H$_2$-H$_2$ \citep{abel_h2-h2, fletcher_h2-h2} and H$_2$-He \citep{abel_h2-he}. Finally, Rayleigh scattering is computed for all possible molecules.   \\ \\

As already mentioned, we automatically couple the planet radius and the planet mass for all 3 regions. For these, we use the parameters from \cite{bonomo_2017}. For this example, we fix the angular size of the terminator region $\theta_K$ to 15$^{\circ}$. In terms of temperature profiles, each region has its own and we do not couple them. We use the n-point model, which presents a convenient way to manipulate T-p relations. This is a purely heuristic profile, where the temperature is linearly interpolated between defined T-p points and has been introduced in the last version of TauREx \citep{al-refaie_taurex3}. We use 5 points to describe the day and the terminator regions, and 3 points for the night side. 

In terms of chemical abundances, we use constant volume mixing ratios with altitude. We couple the molecular profiles from the terminator and the night side. This therefore leaves us with only two parameter per molecule: one for the day side mixing ratio and one for the terminator and night side mixing ratios.  \\ 

All the parameters used for the phase curve forward model and their coupling are described in Table \ref{tab:forward model parameters}.

\begin{table}[h]
\label{tab:forward model parameters}
\centering
\begin{tabular}{lrrr}
\hline\hline \\
Parameters & Day & Terminator  & Night \\
\hline\hline \\
$R_p$ (R$_J$) & 1.036         & coupled     & coupled    \\
$M_p$ (M$_J$) & 2.050         & coupled     & coupled   \\
$T_{surf}$ (K)    & 1850        & 1750     & 500        \\
$T_{1}$ (K)    & 1850        & 1700     & 450       \\
$P_{1}$ (bar)    & $0.2$       & 1    & 1  \\
$T_{2}$ (K)    & 1750        & 1600     & none       \\
$P_{2}$ (bar)    & $6 \times 10^{-2}$     & 0.7    & none  \\
$T_{3}$ (K)    & 1500        & 1250     & none       \\
$P_{3}$ (bar)    & $2 \times 10^{-2}$       & 0.1    & none  \\
$T_{top}$ (K)    & 1450       & 1000    & 400  \\
$P_{top}$ (bar)    & $2 \times 10^{-3}$       & $10^{-2}$    & $10^{-2}$  \\
H$_2$O    & $6 \times 10^{-3}$       & $10^{-5}$    & coupled term  \\
CH$_4$    & $10^{-7}$       & $10^{-4}$    & coupled term  \\
CO    & $10^{-2}$       & $10^{-4}$    & coupled term  \\
\hline\hline \\
\end{tabular}
\caption{Parameters used for the Day, Terminator and Night regions of our WASP-43\,b forward model.}
\end{table}

\begin{center}
\begin{figure}[h]
\centering
    \includegraphics[width=0.48\textwidth]{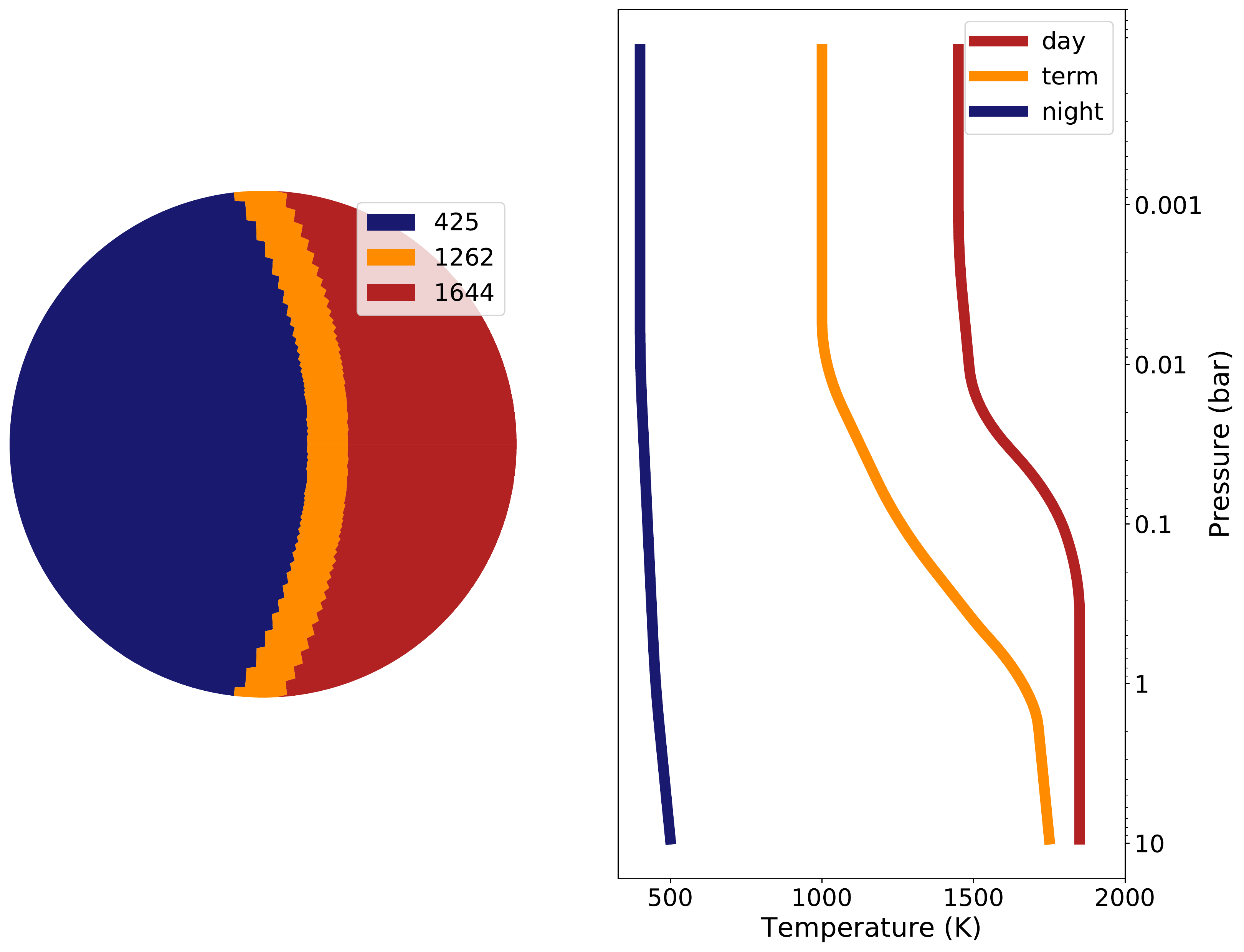}
\caption{Left: Vertically averaged temperature map of our forward model of WASP-43\,b. Right: temperature structure of each region in our phase curve example. These are inspired from the retrieved profiles in \cite{stevenson_w43_2}. We label, In red: day side; In orange: terminator region; In blue: night side.}
\label{fig:phase_temp}
\end{figure}
\end{center}

Figure \ref{fig:phase_temp} shows the temperature profiles and distributions of each region and shares a similar structure to \cite{stevenson_w43_2}.

This setup could be particularly relevant for future atmosphere studies, showing how the complexity of models could be adapted to the information content of each region of the planet. For example, the night side, being more difficult to constrain, would not support a complex chemistry retrieval and temperature retrieval so it would make sense to allow some coupling with the terminator region, which can be more precisely informed by the transit spectrum.

\begin{center}
\begin{figure*}
\centering
    \includegraphics[width=1\textwidth]{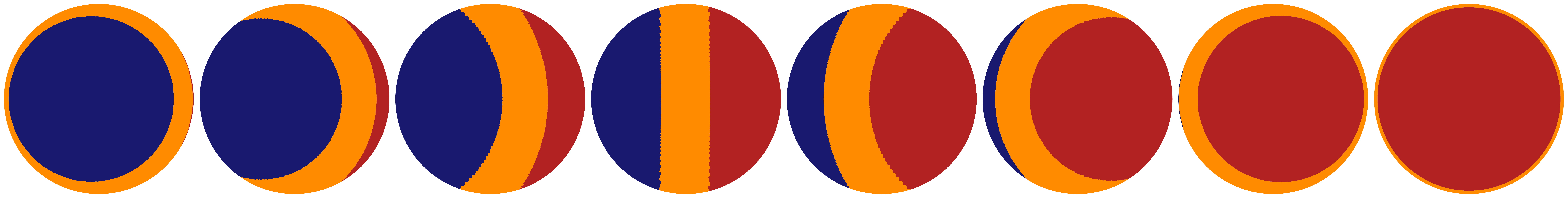}
    \includegraphics[width=1\textwidth]{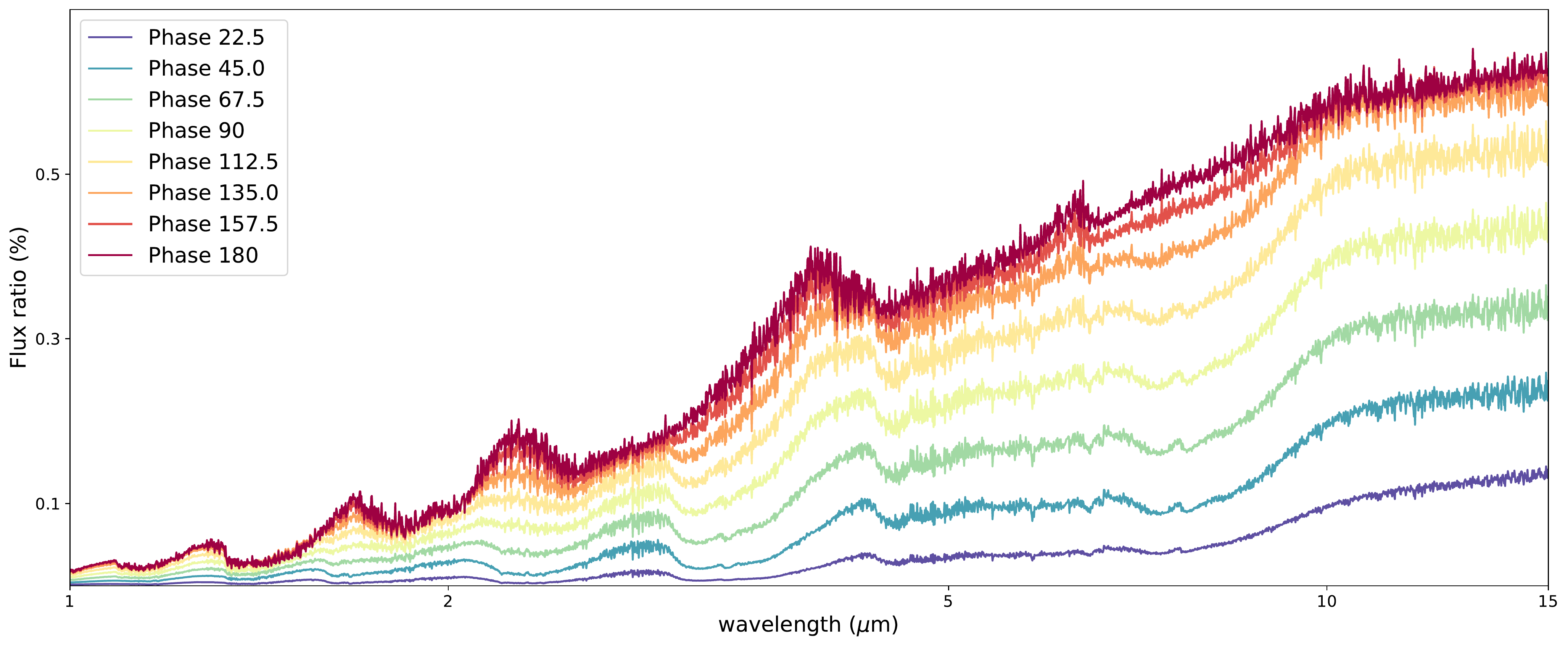}
\caption{Top: Geometry of the phase model at different phases. From left to right: 22.5, 45, 67.5, 90, 112.5, 135, 157.5 and 180. In blue: night side contribution; In orange: terminator contribution; In red: day side contribution. Bottom: Corresponding phase curve emission from our model.}
\label{fig:phase_plots}
\end{figure*}
\end{center}

We run this model for 8 phases with 30 Gaussian quadrature points. The resulting spectra at phases 22.5, 45, 67.5, 90, 112.5, 135, 157.5 and 180 degrees are plotted in Figure \ref{fig:phase_plots}.

%\onecolumngrid
%\twocolumngrid

\begin{center}
\begin{figure}[h]
\centering
    \includegraphics[width=0.45\textwidth]{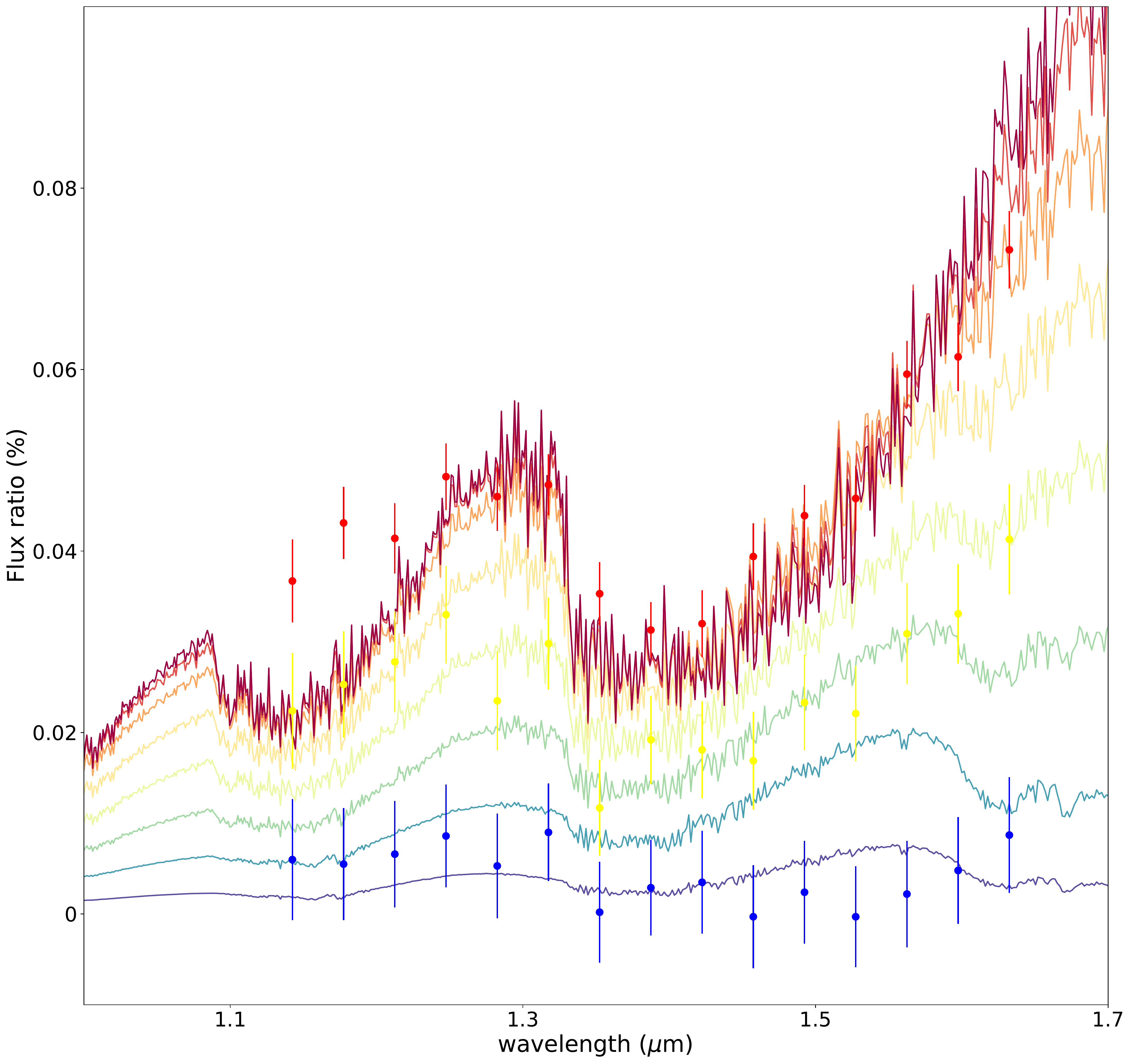}
\caption{Same forward model spectra as in \ref{fig:phase_plots} from our phase curve model of WASP-43\,b. We also plot the HST reduced observations from \cite{stevenson_w43_2} for phase 25 degrees (blue), 90 degrees (yellow) and 180 degrees (red).}
\label{fig:phase_plots_hst}
\end{figure}
\end{center}

In Figure \ref{fig:phase_plots_hst}, we also plot the same model in the Hubble wavelength region and show the observations for phase 25 degrees, 90 degrees and 180 degrees from \cite{stevenson_w43_2}.

As we can see on Figure \ref{fig:phase_plots_hst}, our phase curve forward model is able to reproduce the phase curve observations of WASP-43\,b from the Hubble Space Telescope. Constraining the geometry therefore allows one to limit the number of degrees of freedom, while properly describing the information contained at all phases.

\begin{center}
\begin{figure}[h]
\centering
    \includegraphics[width=0.48\textwidth]{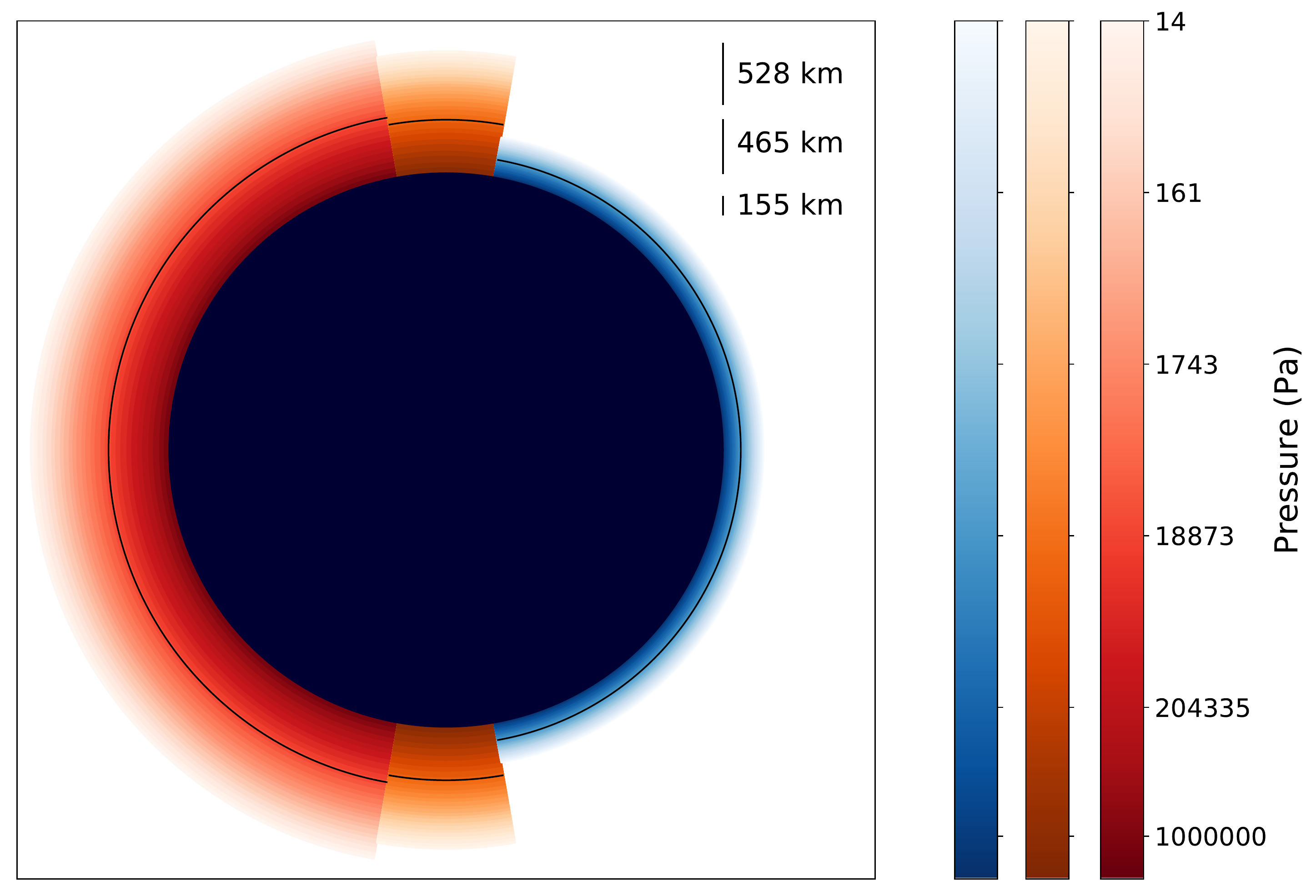}
\caption{Geometry of our phase curve model showing the 3 different altitude pressure profiles: In red: day side; In orange: terminator region; In blue: night side. The strength of the color represents the pressure. For indication, we also show with the black solid line the altitude at $5 H$, where $H$ is the averaged scale height. The corresponding altitude values in kilometers are indicated on the top-right corner.}
\label{fig:phase_alt-press}
\end{figure}
\end{center}

In our phase curve model, the altitude-pressure profile is calculated separately for all regions. This implies that the planet scale height depends on the region, allowing for a better representation of the planet atmospheric structure. Indeed, it has been shown in \cite{caldas_3deffects} that the night side and the day side of tidally locked planets could be very different. We show in Figure \ref{fig:phase_alt-press} the structure of the atmosphere for our WASP-43\,b simulation.

\section{DISCUSSION}

\subsection{Number of Gaussian quadrature points required}

Our numerical integration method requires a fixed number of Gaussian quadrature points. In the literature \citep{Waldmann_taurex2, Irwin_w43b_phase}, eclipse calculations are performed using a small number of Gaussian points (typically less than 10). In this section, we investigate how this parameter impacts the accuracy of our phase curve integration. We assess this by varying the number of Gaussian points in different scenarios and comparing to a reference baseline model with 1000 Gaussian points. In practice, we compare the computed spectra using a single metric $M$:

\begin{equation}
    M = \frac{\sum_{\lambda} \sum_{\Phi} S_{ref}(\lambda, \Phi) \times F(\lambda, \Phi)  }{\sum_{\lambda} \sum_{\Phi}S_{ref}(\lambda, \Phi)},
\end{equation}
where $\lambda$ is the wavelength, $\Phi$ the orbital phase, $S_{ref}$  is the planet to star signal for 1000 Gaussian points and the function F is defined as:
\begin{equation}
    F(\lambda, \Phi) = \frac{| S_{ref}(\lambda, \Phi)-S_{GP}(\lambda, \Phi) |}{S_{ref}(\lambda, \Phi)},
\end{equation}
where $S_{GP}$ is the planet to star signal with a number of Gaussian points to be analysed.

\begin{center}
\begin{figure}[h]
\centering
    \includegraphics[width=0.48\textwidth]{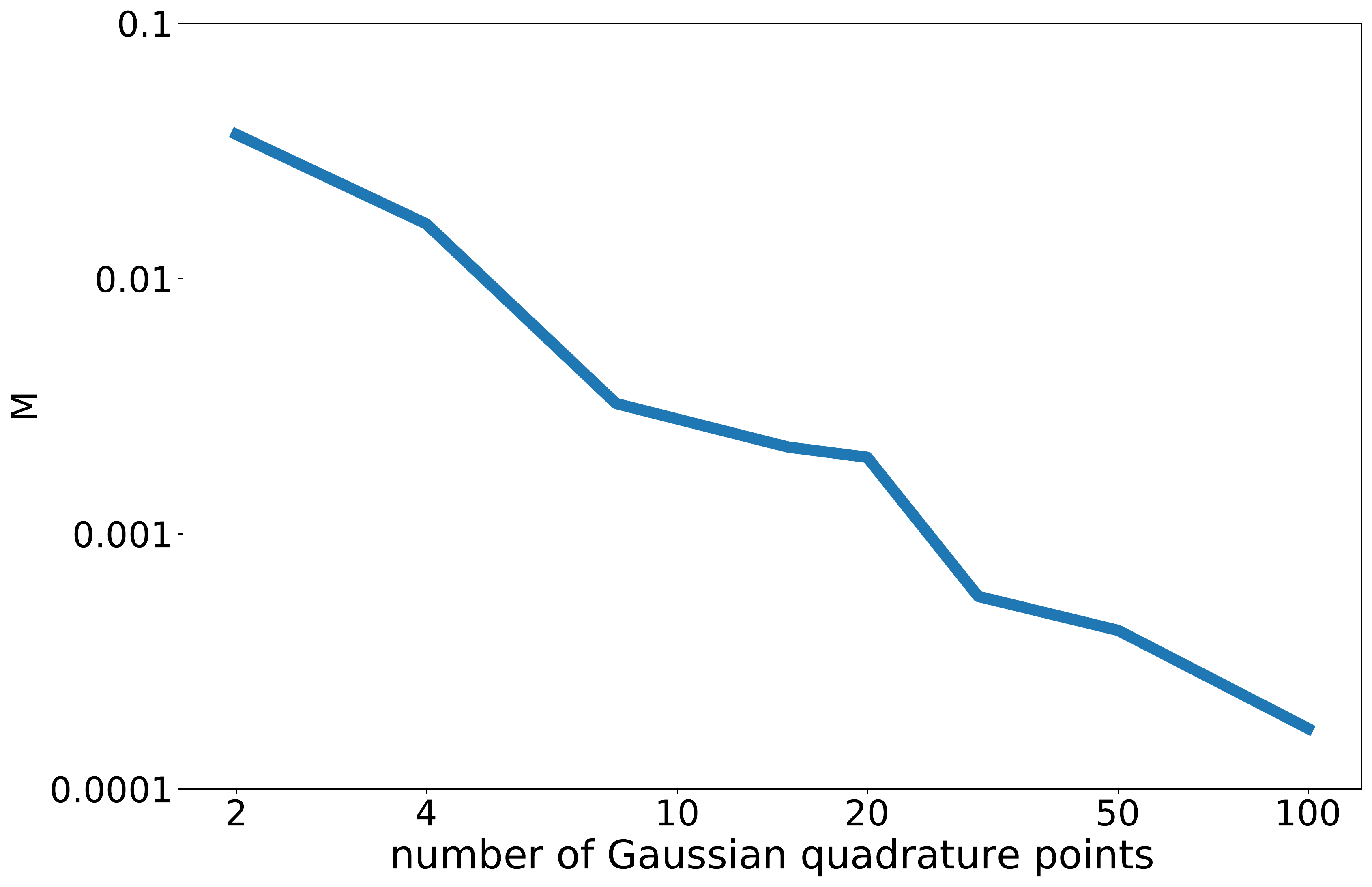}
\caption{$M$ as a function of the number of Gaussian quadrature points in the model. This shows the weighted averaged normalised difference compared to the baseline model with 1000 Gaussian points.}
\label{fig:ngauss_benchmark}
\end{figure}
\end{center}

This represents the weighted average of the normalised distance from the reference model at 1000 Gaussian points, where the weights are the reference model fluxes at each wavelength. In our definition, we use a weighted average to account for the planet flux being lower at small phases, inducing larger but less impacting differences in $| S_{ref}(\lambda, \Phi)-S_{GP}(\lambda, \Phi) |$. We test values of 2, 4, 8, 14, 20, 30, 50 and 100 Gaussian quadrature points. Figure \ref{fig:ngauss_benchmark} shows the normalised difference with the baseline model.

One can see that the accuracy of the model ($M$) scales linearly in log-scale with the number of Gaussian quadrature points. For all Gaussian points, this is lower than the characteristic current noise on phase curve measurements (around 10 percent of the signal in the WASP-43\,b HST spectra presented in this paper). For our applications, we believe that between 10 and 30 Gaussian points represent a good trade-off between accuracy (M is less than 1 percent) and speed.

\subsection{Computational Efficiency}

A single phase calculation requires a minimum of three emission models to successfully complete. We therefore expect $O(N)$ scaling with the number of phase points. This presents a problem, when dealing with multiple phases as we can expect to see a run-time of $ t_{p} = 3Nt_{e}$ where $t_p$ is the time to run our phase model, $N$ is the number of phases and $t_e$ is the time taken to run a single emisson model. This can be circumvented by partially modelling the emission up until the Gaussian-quadrature summation step, then completing the integration for each phase. The heavy calculation is only performed once and each phase only has to perform a much lighter reduction step to produce its flux. We therefore expect a small increase in run-time with each additional phase.

To test this, we use a Macbook-Pro 2017 equipped with a 2.3 GHz Intel Core i5 and we run our phase curve model on a single core. When not specified, we use the same values as for the model presented in the example section. In particular, we have 3 fully separated temperature profiles and 2 sets of 3 molecules, since the terminator and the night side chemistry are coupled. For this example, we use the same cross sections but we limit the calculation to the more common wavelength range of 0.3 $\mu$m - 15 $\mu$m.  Prior to the tests, we run the model once to initialise the profiles and account for preliminary caching steps. This ensures that the time stated refers to the forward model calculation only. We then average the execution time of 4 runs. In our first test, we investigate the impact of calculating the emission at different number of phases simultaneously. This is shown in Table \ref{tab:perf-phase} where we tested the following number of phase points: 1, 10, 100 and 1000. For this test, we use the same number of 30 Gaussian quadrature points.

\begin{table}[]
\centering
\begin{tabular}{lrrrrr}
\hline\hline \\
n Phases  & 1 & 10 & 100 & 1000 & eclipse only \\
\hline \\
Time (s)  &  10.10 &  10.17     & 10.99  & 18.35 & 2.29 \\
\hline\hline  \\
\end{tabular}
\caption{A comparison of the time required to produce a different number of phases with our phase curve forward model (30 Gaussian quadrature points). The simple emission is also shown for comparison.}
\label{tab:perf-phase}
\end{table}

This shows that producing the emission at various phases does not impact much the computing time. Indeed, our previously stated two-step emission solution demonstrates significantly improved scaling compared to a more naive approach with 100 phases only increasing the run-time by 10\%. The coefficients and reduction steps only begin to impact performance at very large numbers of phase calculations. We also confirm that our model is more or less 4 times slower to compute phase curve than it is to compute a standard secondary eclipse emission. This is expected as we are at minimum computing 3 emission models + 1 transmission model every time. \\ \\

The second test concerns the scaling with number of Gaussian quadrature points. We apply the same methodology and we calculate the time required to get 8 simultaneous phases for different number of points. This is shown in Table \ref{tab:perf-gauss}, where we estimated the time for 2, 4, 8, 14, 20, 30, 50 and 100 Gaussian quadrature points.

\begin{table}[]
\centering
\begin{tabular}{lrrrrrrrr}
\hline\hline \\
n Gauss  & 2 & 4 & 8 & 14 & 20 & 30 & 50 & 100 \\
\hline \\
%Time (s)  & 3.771559953689575 &  4.110932767391205     & 4.928640723228455  & 5.944395959377289 & 7.622372806072235 & 10.161592781543732 & 15.16611522436142 & 27.42495948076248, 409.8642784357071 \\
Time (s)  & 3.8 &  4.1 & 4.9 & 5.9 & 7.6 & 10.2 & 15.2 & 27.4 \\
\hline\hline  \\
\end{tabular}
\caption{A comparison of the time required to produce 8 phases with our phase curve forward model for different number of Gaussian quadrature points. }
\label{tab:perf-gauss}
\end{table}

Here, we note that the scaling is much worse. Indeed, the Gaussian points number impacts directly the emission calculation of each model. The emission calculations involve sums and exponential of 2d arrays representing wavelengths and number of layers. \\ \\

The significant computational efficiency demonstrated against a large sample of phases should make this model suitable for standard Bayesian retrieval applications. We attach strong importance to this as it is anticipated that the increased information content from combining phase spectra will require significantly more sampling points (1,000,000+) to reach the necessary evidence tolerance in the retrieval.
\subsection{Limitations of the model}

As seen in the previous section, our phase curve model achieves high performances. To reach this level, we take advantage of the particular geometry to perform the integrals along the phases in a semi-analytical manner. This however means that our phase curve model can only be applied to planets that are compatible with this geometry: the planets must be tidally locked or in spin-synchronous orbits, for which the regions can be approximated by homogeneous temperature and chemical structure and/or for which the available data is not detailed enough to support a more granular model. In our model we only resolve 3 regions, in some cases for the next generation of space telescopes such as ESA-Ariel \citep{Tinetti_ariel}, NASA-JWST \citep{Bean_JWST} or Twinkle \citep{Edwards_twinkle} it may be necessary to push to more detailed schemes with more than 3 regions or to a continuous description of the geometry. Thanks to a recent rework, the new architecture of TauREx 3 is now very flexible and easily modifiable, which means that the work presented in this paper could be rapidly extended. Other limitations include the plane-parallel assumptions made in Equation \ref{eq:base_em}. While each region possess its own scale height, the planet curvature leads to terminator emission through more complicated atmospheric paths at phase angles close to 180$^{\circ}$. These effects are not accounted in our model as they would require a full 3-dimensional treatment \citep{caldas_3deffects}. Other effects described in \cite{caldas_3deffects} or \cite{MacDonald_2020}, such as the transmission through multiple atmospheric regions in transit scenarios or the differences between morning and evening terminator, could in theory be implemented with the family of models presented here.

\subsection{Retrieval possibilities and advantage}

As shown previously in the discussion, the support of our phase curve models does not bring huge performance losses compared to our standard forward model. This means that potentially, this model could be improved to be used in a retrieval setting. Indeed, this description, which in essence only combines simpler emission and transmission models in a higher hierarchical model, would be convenient as it is fully compatible with the other available modules in TauREx 3 and it already supports the coupling of parameters. As shown in \cite{Irwin_w43b_phase} for the planet WASP-43\,b, a retrieval combining spectra at different phases in a single model allows the efficient recovery of the information content in the dataset by handling the redundant information in a unified way.

\section{CONCLUSIONS}

Using the flexibility of the next generation of the TauREx retrieval framework \citep{al-refaie_taurex3}, we have constructed a new analytical phase curve model. We describe the planet geometry using 3 distinct region and allow for full control of these regions through parameter coupling (such as radius and mass, or user dependant) to consider the planet as a whole. The forward model calculation is handled through an analytical formulation of the phase geometry, which we combine with the standard emission model of TauREx. This new approach ensures a very fast computation time (only 4 times slower than a single emission model), which only weakly scales with the number of phases to simulate. In the future, we intend to test this model further on real case scenarios and investigate potential improvements which could be made to prepare for the next generation of space telescopes.

\vspace{5mm}
\noindent\textbf{Acknowledgements}

This project has received funding from the European Research Council (ERC) under the European Union's Horizon 2020 research and innovation programme (grant agreement No 758892, ExoAI) and under the European Union's Seventh Framework Programme (FP7/2007-2013)/ ERC grant agreement numbers 617119 (ExoLights). Furthermore, we acknowledge funding by the Science and Technology Funding Council (STFC) grants: ST/K502406/1, ST/P000282/1, ST/P002153/1 and ST/S002634/1.

We acknowledge the availability and support from the High Performance Computing platforms (HPC) DIRAC and OzSTAR, which provided the computing resources necessary to perform this work.

We wish to thanks the referee for his/her great suggestions and the relevance of his/her comments.


%%%%%%%%%%%%%%%%%%%%%%%%%%%%%%%%%%%%%%%%%%%%%%%%%%%%%%%%%%%%%%%%%%%%%%%%%%%%%%%%

%%%%%%%%%%%%%%%%%%%%%%%%%%%%%%%%%%%%%%%%%%%%%%%%%%%%%%%%%%%%%%%%%%%%%%%%%%%%%%%%

%\onecolumngrid
\newpage
\bibliographystyle{aasjournal}
\bibliography{main}
%\twocolumngrid

%%%%%%%%%%%%%%%%%%%%%%%%%%%%%%%%%%%%%%%%%%%%%%%%%%%%%%%%%%%%%%%%%%%%%%%%%%%%%%%%

\renewcommand{\floatpagefraction}{.9}%
\section{APPENDIX}

\subsection*{Appendix 1: Derivation of the phase integration coefficients}

Let's consider the situation presented in Figure \ref{fig:geometry}. For this derivation, we normalise the problem and describe the planet as a sphere (or a circle in 2 dimension) of radius 1. We define the orthonormal basis ({\bf e$_x$}, {\bf e$_y$}) associated with coordinate (x,y) and the corresponding polar coordinates ($r$, $\alpha$). \\ 

In our model, the mean terminator is described by a circle of centre (x$_0$, y$_0$) and radius R. As it must pass through the points of coordinate (0, 1), (0,-1) and (cos($\Phi$),0), where $\Phi$ is the phase angle (angle observer-star-planet), we immediately get the terminator equation:
\begin{equation}\label{eq:circle_coord}
    (x-x_0)^2 + y^2 = R^2
\end{equation}
with:
\begin{equation}
    x_0 = \frac{ \mathrm{cos}(\Phi)^2 - 1}{2 \mathrm{cos}(\Phi)},
\end{equation}
and:
\begin{equation}
    R^2 = x_0^2+1.
\end{equation}

This equation is valid for a terminator region of size 0. For a terminator region of angular size $\theta_K$, where $\theta_K$ is the spherical angle between the two boundaries of the terminator, we consider the representation shown in Figure \ref{fig:geometry}. As $\theta_K$ is defined on the sphere, it is linked to the projected distance $K_{\pm}$ from the terminator centre to the terminator boundaries by: 
\begin{equation}
    K_{\pm} = |\mathrm{cos}(\Phi) - \mathrm{cos}(\Phi \pm \theta_K)| 
\end{equation}

In this case, the boundaries of the terminator region are described by the same  Equation \ref{eq:circle_coord}, with only a change in the radius of the terminator circle ($R' = R \pm K_{\pm}$). We get the following equation:

\begin{equation}
    (x-x_0)^2 + y^2 = \left(\sqrt{x_0^2+1} \pm K_{\pm} \right)^2.
\end{equation}

Developing this equation and shifting to the polar coordinates $x = r \mathrm{cos}(\alpha)$ and $y = r \mathrm{sin}(\alpha)$ leads to:

\begin{equation}
    r^2 - 2 x_0 r  \mathrm{cos}(\alpha) = 1 + K_{\pm}^2 \pm \sqrt{x_0^2+1}.
\end{equation}\label{eq:interm}

Now as we are looking for the intersection point between our terminator boundaries and the integration circle of radius sin$(\theta) = \sqrt{1-\mu^2}$, we can add the additional constraint of:

\begin{equation}
    r^2 = 1 - \mu^2.
\end{equation}

Plugging this in Equation \ref{eq:interm} leads to the desired relation:

\begin{equation}
    \mathrm{cos}(\alpha) =  \frac{\mathrm{cos}(\Phi)}{ (1-\mathrm{cos}^2(\Phi))\sqrt{1-\mu^2}}\left( \mu^2 + K_{\pm}^2 \pm 2 K_{\pm} \sqrt{1 +\frac{(\mathrm{cos}^2(\Phi)-1)^2}{4 \mathrm{cos}^2(\Phi)}} \right). 
\end{equation}

We note that this equation is not defined for $\mu = 1$ as, in this case, the integration circle corresponds to a unique point. Similarly, in the case of $\Phi = 90$ exactly, the terminator boundaries are not defined by circles anymore but by vertical lines. This situation require a separated treatment and, using the same approach, we find the simplified form:

\begin{equation}
    \mathrm{cos} (\alpha ) =  \frac{\mathrm{sin}(\theta_K)}{\sqrt{1-\mu^2}}.
\end{equation}

\newpage
\subsection*{Appendix 2: Derivation of the relation between the phase angle $\Phi$ and the time t.}

Equation \ref{eq:phase_eq} provides $\Phi$ the angle between the observer, the star and the planet. If the planet orbit is circular, it can be mapped to the time t easily using a linear mapping. The following formula give the phase angle for the circular case $\Phi_{cir}(t)$:
\begin{equation}
    \Phi_{cir}(t) = 2\pi \frac{t}{T}, 
\end{equation}
where T is the orbital period of the planet. \\

In the case of non-circular orbits, one must solve the Kepler's equations to adapt this mapping. For the tidally locked case, we label this new angle $\Phi_{tid}(t)$, which can be derived from the classical Kepler's laws. Here, we reproduce and adapt the classical derivations following \cite{1993_colwell,sertorio2001constraints,dvorak_extrasolar_2008, seager_exoplanets_2010, lissauer_fundamental_2013, perryman_exoplanet_2018}.\\

\begin{figure*}
\begin{center}
\begin{tikzpicture}

\draw [ultra thick, dashed] (0,0) circle (5);
\draw [ultra thick](0,0) ellipse (5 and 3);

\draw [color=black,dashed, ->] (0,0) -- (5.5,0);
\node (a) at (5.9, 0) {$x$};
\draw [color=black,dashed, ->] (0,0) -- (0,5.5);
\node (b) at (0, 5.9) {$y$};

\draw [color=gray, dashed] (0,0) -- (3.6,2);
%\draw [color=blue, dashed] (3.6,2) -- (4.4,2.5);

\draw [color=gray, dashed] (3.6,0) -- (3.6,3.5);

\draw [color=gray, dashed, ultra thick] (0,0) -- (3.6,3.5);

\draw [color=red, ultra thick] (2.5,0) -- (3.6,2);
\draw [color=blue, ultra thick] (2.5,0) -- (3.6,3.5);
\draw [red,  ultra thick] (4.9,0) arc(0:42:4.9 and 2.9);
\draw [blue,  ultra thick] (4.9,0) arc(0:44:4.9 and 4.9);
\draw [color=red,  ultra thick] (2.5,0.05) -- (5,0.05);
\draw [color=blue,  ultra thick] (2.5,0.1) -- (5,0.1);

\fill [yellow] (2.5,0) circle[radius=0.3];
\fill [brown] (3.65,2) circle[radius=0.3];
\node [black] (f) at (2.5, 0) {S};
\node [white] (g) at (3.65, 2) {P};

\draw [orange,  ultra thick, <->] (1,0) arc(0:44:1 and 1);
\node (b) at (1.1, 0.5) {E};
\draw [orange,  ultra thick, <->] (3,0) arc(0:18:3 and 2);
\node (c) at (3.5, 0.4) {$\alpha - \omega$};
\draw [orange, dashed,  ultra thick, <->] (-0.05,0.25) -- (3.35,3.63);
\node (d) at (1.4, 2) {a};
\draw [orange, dashed,  ultra thick, <->] (0,-0.4) -- (2.5,-0.4);
\node (e) at (1.4, -0.7) {ae};

\end{tikzpicture}
\end{center}
\caption{Illustration of the 2 dimensional trajectory for a planet in an eccentric orbit (e = 0.8). The planet (brown P node) is orbiting the star (yellow S node) following the solid black ellipsis from a position $t_0$ (y = 0) to a position $t$. In dashed black we show the circle of radius corresponding to the semi-major of the ellipsis a. From there, one can construct the angle E as the angle between the x axis and the line from the origin to the planet projection directed by $\textbf{e}_y$ onto the circle. The blue and red areas are the areas of interest for our problem.} \label{fig:orbit}
\end{figure*}
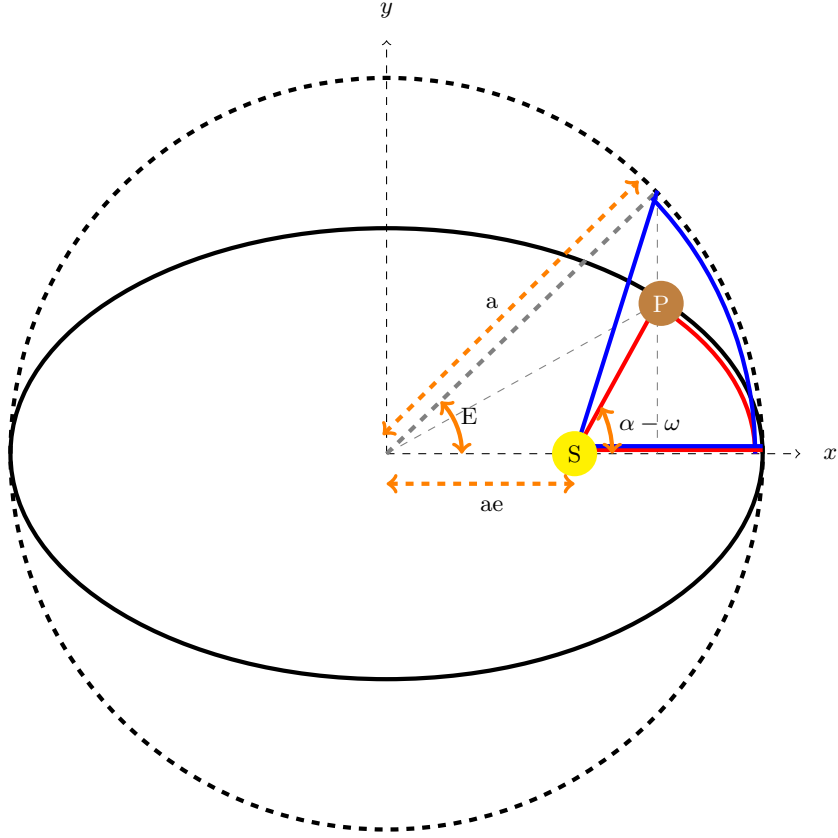

The Equation Of Motion (EOM) for a central gravitational force is given by \cite{Newton}:
\begin{equation}
    \Ddot{\mathbf{r}} + \frac{\mathrm{GM}_{s} \mathbf{r}}{r^2} = 0,
\end{equation}
where $\Ddot{\mathbf{r}}$ refers to the second time derivative of $\mathbf{r}$ (the 'dot' notation means time derivative), G is the gravitation constant and M$_s$ is the stellar mass. $\mathbf{r}$ (in 'bold') refers to the vector of magnitude r, from the planet towards the centre of the star (axis $\mathbf{e_r}$). The vectors $\mathbf{r}$ and $\Ddot{\mathbf{r}}$ can be expressed in polar coordinates ($r$, $\alpha$) as:

\begin{equation}
\begin{split}
    &\mathbf{r} = r \mathbf{e_r}, \\
    &\dot{\mathbf{r}} = \dot{r} \mathbf{e_r} + r \dot{\alpha} \mathbf{e_{\alpha}}, \\
    &\Ddot{\mathbf{r}} = (\Ddot{r} - r \dot{\alpha}^2)\mathbf{e_r} + \frac{1}{r} \dv[]{t}\left(r^2\dot{\alpha} \right) \mathbf{e_{\alpha}}.
\end{split}
\end{equation}
The projection of the EOM on $\mathbf{e_{\alpha}}$ allows us to recover the angular momentum constant L: 
\begin{equation}
    L = r^2\dot{\alpha} = \mathrm{constant}
\end{equation}
For the projection on the $\mathbf{e_r}$ axis, we apply the change of variable  u = 1/r. We therefore have:
\begin{equation}
\begin{split}
    &\dot{u} = - u^2 \dot{r}, \\
    &\Ddot{u} = -2 \dot{u} u \dot{r} - u^2 \Ddot{r}.
\end{split}
\end{equation}
Using L and noting that $\dv{t} = \dot{\alpha} \dv{\alpha}$ the OEM on $\mathbf{e_r}$ transforms into:
\begin{equation}
    u'' - u = \frac{-GM_s}{L^2}.
\end{equation}
Where the 'prime' notation refers to the derivative with $\alpha$. This classical second order differential equation is known as the Binet's equation and has solutions of the general form:
\begin{equation}
    u = \frac{-GM_s}{L^2}(1+A \mathrm{cos}(\alpha - B)),
\end{equation}
where A and B are constants depending on the initial conditions. Using the classical definitions of the semi-major axis a, the eccentricity e and the longitude of the pericentre $\omega_0$, the final solution for r is:
\begin{equation}
    r(\alpha) = \frac{a(1-e^2)}{1+e \mathrm{cos}(f)},
\end{equation}
where we define $f = \alpha - \omega_0$, the true anomaly.
These solutions have the forms of ellipsis, hyperbola or parabola. For our example, we consider gravitationally bounded orbits so the solutions will take the form of an ellipsis, with e $<$ 1. This is shown in Figure \ref{fig:orbit}, where the eccentricity is 0.8 .  \\

In this formulation, we unfortunately eliminated the time t. As we want to express $\Phi(t)$ we need to transform our solution to express $r(t)$ and $\alpha(t)$.

This can be done by defining the mean anomaly M as the angular distance to the pericentre:
\begin{equation}
    M = \frac{2 \pi}{T}\left(t - t_0 \right).
\end{equation}
M does not have a physically evident interpretation but it is related to an angle called the eccentric anomaly E (see Figure \ref{fig:orbit}).

Thanks to Kepler's law of equal areas, we also have the relationship:
\begin{equation}
    M = 2\pi \frac{A(t)}{A_{total}},
\end{equation}
where $A$ designs the surface of the ellipsis that is cut during a specific time t. $A_{total}$ is the entire surface of the ellipsis and is equal to $\pi a b$.

In Figure \ref{fig:orbit}, A(t) corresponds to the red area. It is related to the blue area A'(t) by the relation A(t) = b/a A'(t). From Figure \ref{fig:orbit}, we can express A'(t) as:
\begin{equation}
    A'(t) = \frac{1}{2}a^2 E(t) - \frac{1}{2} a^2 e \mathrm{sin}(E).
\end{equation}
Finally, we find the Kepler's equation:
\begin{equation}\label{eq:kepler}
    \frac{2\pi}{T} (t-t_0) = E - e \mathrm{sin}(E).
\end{equation}

Now the eccentric anomaly E can be related to $r$ and $\alpha$ by noting that:
\begin{equation}
\begin{split}
    &x = a \mathrm{cos}(E), \\
    &x = r \mathrm{cos}(f) + a e.
\end{split}
\end{equation}
This leads to:
\begin{equation}
\begin{split}
    &r = a(1-e \mathrm{cos}(E)), \\
    &\mathrm{cos}(f) = \frac{\mathrm{cos}(E)-e}{1 - e \mathrm{cos}(E)}.
\end{split}
\end{equation}
So, provided we can solve Equation \ref{eq:kepler} for E, we can now link the time t with the angular position of the planet $\alpha$. The Kepler's equation can't be solved directly but multiple numerical or iterative procedures exist \citep{1979_smith,1983_danby,1989_taff_kepler,1993_colwell,2000_Murray_dynamics,Boyd2013FindingTZ}. In this work, we use the following iterative scheme:
\begin{equation}
\begin{split}
    &E_0 = M, \\
    &E_{i+1} = M + e \mathrm{cos}(E_i).
\end{split}
\end{equation}

\begin{figure*}\label{fig:geom_3d}
\begin{center}
\begin{tikzpicture}

%\draw [ultra thick, dashed] (0,0) circle (5);
\draw [](0,0) ellipse (5 and 2);
\draw[ultra thick, rotate around={-30:(0,0)},black](0,0) ellipse (4.5 and 1.5);
\draw[rotate around={-30:(0,0)},black, dashed] (0,0) -- (0,2);
\draw[rotate around={150:(0,0)},black, dashed] (-5,0) -- (5,0);
    
\draw[rotate around={18:(0,0)},dashed] (0,-3.2) -- (0,3.2);
\draw[blue,rotate around={18:(0,0)},->] (0,0) -- (0,-2.8);
\draw[blue, ->] (0,0) -- (0,3);
\draw[blue, ->] (0,0) -- (5.5,0);

\draw [orange,  ultra thick, <->] (-0.7,-2) arc(178.5:163:3 and 2);
\node (a) at (-0.9, -1.7) {$I$};

\draw [orange,  ultra thick, <->] (-0.5,0.3) arc(170:250:1 and 0.9);
\node (o) at (-0.8, -0.3) {$\omega$-$\pi$};

\node (p) at (-4.3, 2.8) {Periastron};
\node (n1) at (-0.5, 2.3) {N1};
\node (n2) at (0.2, -2.2) {N2};

\node (x) at (1.2, -2.9) {{\bf X}};
\node (y) at (5.8, 0) {{\bf Y}};
\node (z) at (0, 3.2) {{\bf Z}};

\fill [yellow] (0,0) circle[radius=0.3];
\fill [brown] (-3,0.4) circle[radius=0.3];
\node [black] (f) at (0, 0) {S};
\node [white] (g) at (-3,0.4) {P};

\end{tikzpicture}
\end{center}
\caption{3-dimensional representation of a planet along its orbit. The planet trajectory (in bold) forms an ellipsis with inclination angle I from the celestial sphere while N1 and N2 represents the two nodes. The Z axis is along the observer line of sight. } 
\end{figure*}
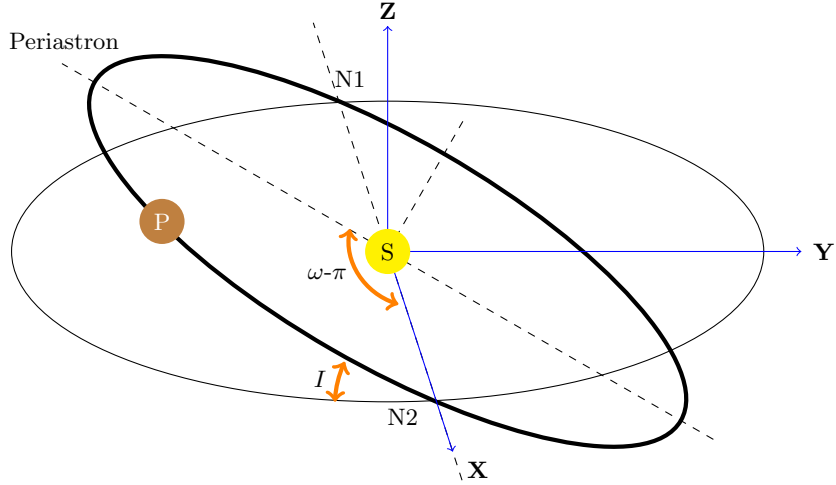

This series is convergent and converges towards E. Now, the final remaining step is to relate the angle $\alpha$ to the angle observer-star-planet $\Phi$. This can be done using the standard rotation matrices to transform the local ellipsis coordinate system (x,y) into any generic coordinate system (X,Y,Z):
\begin{equation}
    \begin{pmatrix}
    X \\ Y \\ Z
    \end{pmatrix}
    = R_I R_{\omega} R_{\Omega}
    \begin{pmatrix}
    x \\ y \\ z
    \end{pmatrix}
\end{equation}
where the introduced R matrices describe the rotations for the argument of the pericentre $\omega = \Omega +\omega_0$, the inclination I and the longitude of the ascending node $\Omega$. 

\begin{equation}
    R_{I} = \begin{pmatrix}
        1 & 0 & 0 \\
        0 & \mathrm{cos}(I) &  -\mathrm{sin}(I) \\
        0  & \mathrm{sin}(I)  &  \mathrm{cos}(I) 
\end{pmatrix}
\end{equation}
\begin{equation}
    R_{\omega} = \begin{pmatrix}
        \mathrm{cos}(\omega-\Omega) & -\mathrm{sin}(\omega-\Omega) & 0 \\
        \mathrm{sin}(\omega-\Omega) & \mathrm{cos}(\omega-\Omega) &  0 \\
        0  & 0  &  1 
\end{pmatrix}
\end{equation}
\begin{equation}
    R_{\Omega} = \begin{pmatrix}
        \mathrm{cos}(\Omega) & -\mathrm{sin}(\Omega) & 0 \\
        \mathrm{sin}(\Omega) & \mathrm{cos}(\Omega) &  0 \\
        0  & 0  &  1 
\end{pmatrix}
\end{equation}
This definition allows to express the planet trajectory in a general coordinate system, however, in the field of exoplanets, it is common to have the observer on the Z axis and fix $\Omega = \pi$. The X and Y axis, then remain on the plane that is perpendicular to the line of sight with the X axis oriented along the orbit nodes. A schematic of this geometry is presented in Figure \ref{fig:geom_3d}.

Finally, the coordinates T for planet trajectory along its orbit can be expressed in this 3D coordinate system:

\begin{equation}
    T = \begin{pmatrix}
T_X \\ T_Y \\ T_Z
\end{pmatrix} = 
\begin{pmatrix}
    \mathrm{cos}(\omega) & -\mathrm{sin}(\omega)  & 0 \\
    \mathrm{sin}(\omega) \mathrm{cos}(I) & \mathrm{cos}(\omega)\mathrm{cos}(I) & \mathrm{sin}(I) \\
    -\mathrm{sin}(\omega) \mathrm{sin}(I) & -\mathrm{cos}(\omega) \mathrm{sin}(I) & \mathrm{cos}(I)
\end{pmatrix}
\begin{pmatrix}
r \mathrm{cos}(f) \\ r \mathrm{sin}(f) \\ 0
\end{pmatrix}
\end{equation}

%\begin{equation}
%T = \begin{pmatrix}
%T_X \\ T_Y \\ T_Z
%\end{pmatrix} = 
%\begin{pmatrix}
%    \mathrm{cos}(\Omega) \mathrm{cos}(\alpha-\Omega) - \mathrm{sin}(\Omega)\mathrm{sin}(\alpha-\Omega)\mathrm{cos}(I) \\
%    \mathrm{sin}(\Omega) \mathrm{cos}(\alpha-\Omega) + \mathrm{cos}(\Omega)\mathrm{sin}(\alpha-\Omega)\mathrm{cos}(I) \\
%    \mathrm{sin}(\alpha-\Omega)\mathrm{sin}(I)
%\end{pmatrix}
%\end{equation}

As in (X,Y,Z) the direction of reference for the observer is along the Z axis. We can then get the final phase angle for tidally locked planets $\Phi_{tid}(t)$:
\begin{equation}
    \mathrm{cos}(\Phi_{tid}(t)) = \frac{-T_Z}{||T||}
\end{equation}

\newpage

\subsection*{Appendix 3: Examples of planet trajectories}
\begin{center}
\begin{figure}[h]
\centering
    \includegraphics[width=0.88\textwidth]{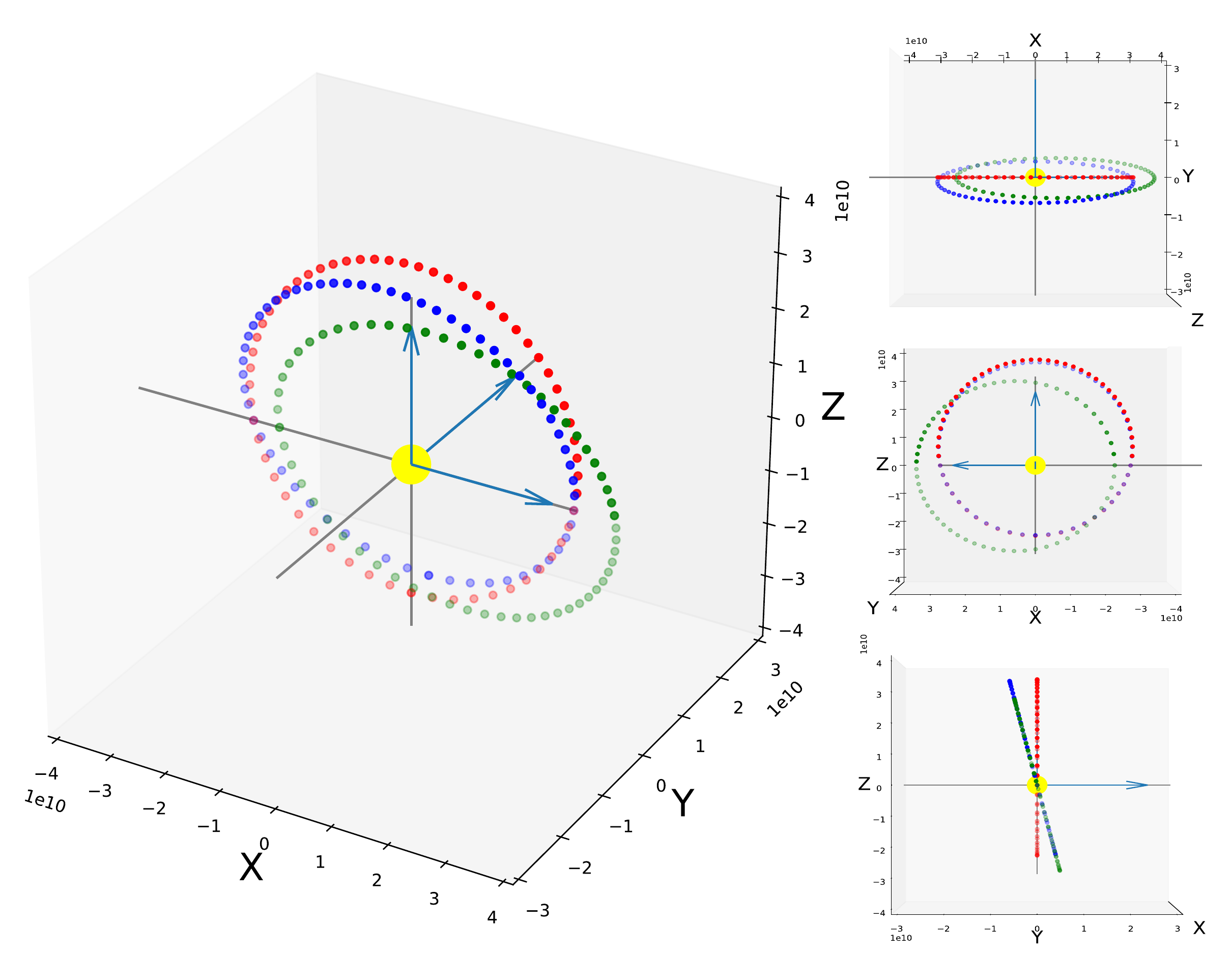}
    \includegraphics[width=0.88\textwidth]{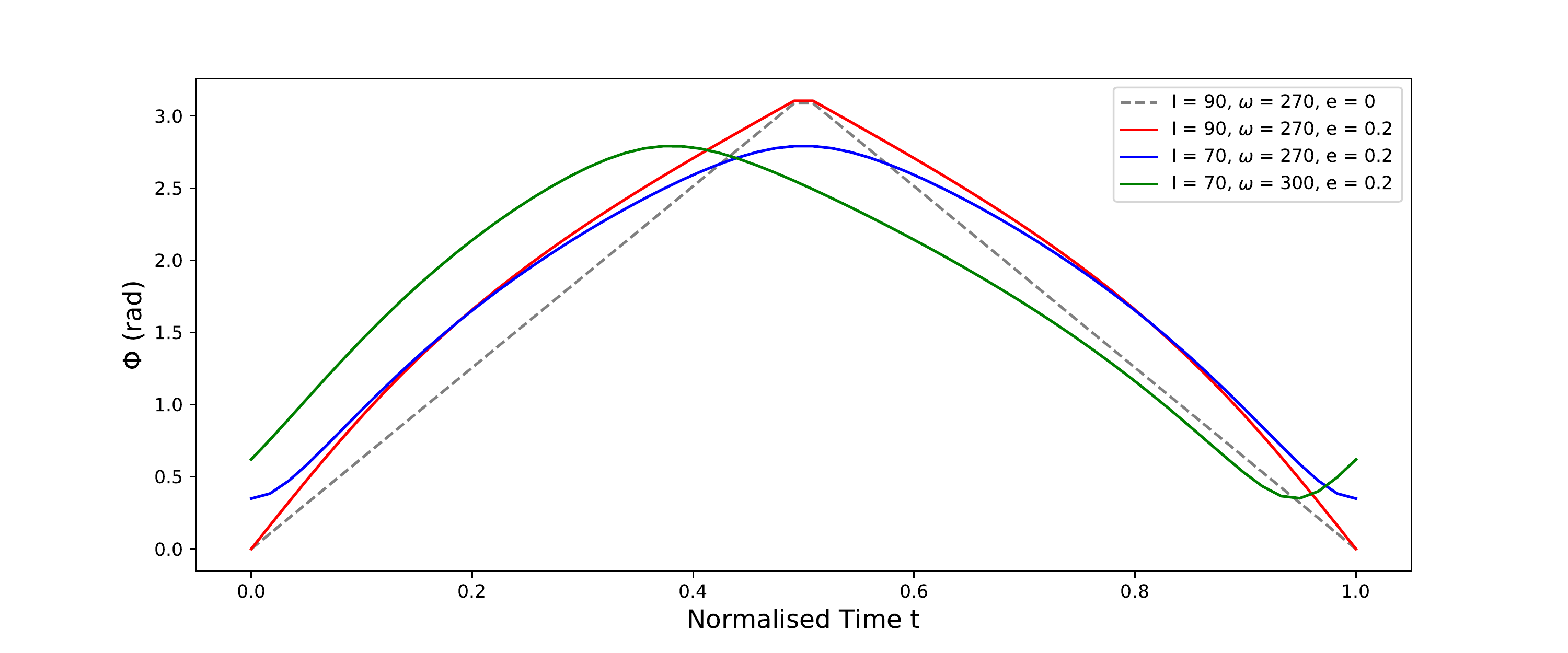}
\caption{Examples of planet trajectories for 3 cases. We show a planet position at 60 different times t in elliptic orbits (a = 0.2 AU and e = 0.2) for: I=90$^{\circ}$ and $\Omega$=0$^{\circ}$ (red); I=45$^{\circ}$ and $\Omega$=0$^{\circ}$ (blue); I=45$^{\circ}$ and $\Omega$=30$^{\circ}$ (green). The top panel shows the 3-dimensional trajectory (left) and the projections to the (X,Y), (X,Z) and (Y,Z) planes  (right, respectively from top to bottom). The corresponding values for the angle $\Phi$ are displayed in the bottom panel. We also add a circular case with e=0 for reference in grey dashed lines. }
\label{fig:trajectory}
\end{figure}
\end{center}

 %%% END BF FOR THE REVIEWER
%\appendix 

%\section*{Appendix: if needed}\label{Appendix_parameters}

%Appendix here if needed 

%%%%%%%%%%%%%%%%%%%%%%%%%%%%%%%%%%%%%%%%%%%%%%%%%%%%%%%%%%%%%%%%%%%%%%%%%%%%%%%%
\end{document}